\documentclass[12pt]{article}
\usepackage{enumerate}
\usepackage[numbers]{natbib}
\usepackage{url} 
\usepackage{enumitem}
\usepackage{amsmath,amssymb,amsthm}
\usepackage{xcolor} 
\definecolor{myblue}{RGB}{0, 0, 188}
\definecolor{mygreen}{RGB}{0, 188, 0}
\usepackage[colorlinks=false]{hyperref}
\usepackage{algorithm,algorithmic}
\usepackage{graphicx,multirow}
\usepackage{xr} 
\newcommand{\blind}{1}

\newtheorem{theorem}{Theorem}
 
\newtheorem{proposition}{Proposition} 
\newtheorem{remark}{Remark}
\newtheorem{corollary}{Corollary}

\newtheorem{assumption}{Assumption}

\begin{document}

	\def\spacingset#1{\renewcommand{\baselinestretch}%
		{#1}\small\normalsize} \spacingset{1}

	
	\if1\blind
	{
		\title{\bf Individualized Dynamic Mediation Analysis Using Latent Factor Models}
		\author{Yijiao Zhang$^{1}$, Yubai Yuan$^{2}$\footnote{Corresponding author. Email address: yvy5509@psu.edu}, Yuexia Zhang$^{3}$, Zhongyi Zhu$^{4}$, Annie Qu$^{5}$ \hspace{.2cm}\\
			$^{1}$Department of Biostatistics, Epidemiology and Informatics,\\ 
       University of Pennsylvania\\
			$^{2}$Department of Statistics, The Pennsylvania State University\\
			$^{3}$Department of Statistics and Data Science,\\ 
			The University of Texas at San Antonio \\
            $^{4}$Department of Statistics and Data Science, 
       Fudan University\\
			$^{5}$Department of Statistics and Applied Probability, \\University of California, Santa Barbara}
		\date{}
		\maketitle
	} \fi
	
	\if0\blind
	{
		\bigskip
		\bigskip
		\bigskip
		\begin{center}
			\setlength{\baselineskip}{1.5\baselineskip}
			{\LARGE\bf  Individualized Dynamic Mediation Analysis Using Latent Factor Models}
		\end{center}
		\medskip
	} \fi
	
	\bigskip
	\begin{abstract}
		Mediation analysis plays a crucial role in causal inference as it can investigate the pathways through which treatment influences outcome. Most existing mediation analysis assumes that mediation effects are static and homogeneous within populations. However, mediation effects usually change over time and exhibit significant heterogeneity among individuals in many real-world applications. Additionally, the mediation mechanism can be complicated and involves non-sparse, making mediator selection particularly challenging. To address these issues, we propose an individualized dynamic mediation analysis method for mediator selection. Our approach can identify the significant mediators at the population level while capturing the time-varying and heterogeneous mediation effects at the individual level via varying-coefficient structural equation models. Another advantage of our method is that we allow the presence of unmeasured time-varying confounders that induce the heterogeneous mediation effects. We provide asymptotic results for the proposed estimator and selection consistency for significant mediators. Extensive simulation studies and an application to a DNA methylation study demonstrate the effectiveness and advantages of our method.
	\end{abstract}
	
	\noindent%
	\vfill
	
	\newpage
	\spacingset{1.9} 
	\section{Introduction}
	
	Mediation analysis is an important tool for investigating the causal pathway from a treatment to an outcome in the presence of intermediate variables, or mediators. It provides mechanistic insights beyond traditional causal inference \citep{NBERt0118} and is essential for designing targeted interventions in many epidemiological studies \citep{bhootra2023dna}.
	
	Mediation could be a dynamic process. For example, in epigenetic studies, DNA methylation (DNA-m) has been highlighted as a potential mediator of the effect of environmental exposures on various diseases  \citep{rusiecki2013ptsd,fransquet2018blood}, and research has suggested that mediation effects through DNA-m change over time \citep{xu2017emerging}. To analyze dynamic mediation effects using longitudinal data, there are two major challenges as follows.

	The primary challenge lies in \textit{individual heterogeneity}. Mediation effects can vary substantially across individuals, as observed in epigenomic studies \citep{castellini2024molecular}. Such heterogeneity arises from genetic, demographic, and environmental differences as well as unobserved individual characteristics. Furthermore, in longitudinal studies, the dynamic mediation effects may also vary across individuals. Therefore, ignoring individual heterogeneity could lead to biases or misinterpretations. 

    Traditional mediation analysis mainly focuses on homogeneous models. For cross-sectional data, notable foundational works include \cite{imai2010identification,imai2013identification,vanderweele2014effect,vanderweele2017mediation}.
	In the context of longitudinal data, \cite{vanderweele2017mediation} establishes non-parametric identification results using a mediation g-formula, accommodating time-varying exposures and mediators. More recently, \cite{cai2022estimation} proposed a varying-coefficient mediation model employing local polynomial regression for intensive longitudinal data. However, these methods have limited capacity to address heterogeneity in mediation mechanisms.

\begin{figure}[H]
    \centering
    \includegraphics[width=0.9\linewidth]{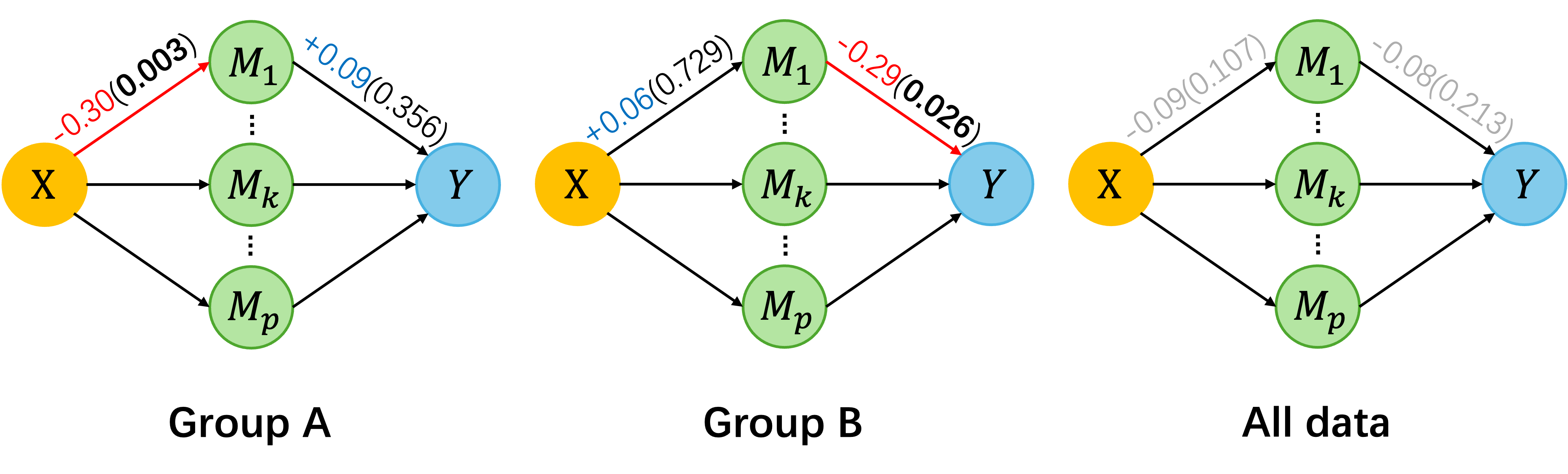}
    \caption{Estimated effects for mediator $M_1$ (cg12682382) in two latent groups with sample sizes larger than 50 and the combined data. Numbers in parentheses represent the regression $p$-values for testing each effect against zero. $X$ denotes depression level and $Y$ denotes cognitive dysfunction. The results are based on the data in year 2010.}
    \label{fig:moti-ex}
\end{figure}

    As a motivating example, we analyze data from the Alzheimer’s Disease Neuroimaging Initiative (ADNI), a longitudinal study of Alzheimer’s progression. We apply our method to investigate how geriatric depression (GDS) influences disease progression through DNA methylation changes at specific CpG sites. Our method identified latent subgroups with distinct mediation patterns. We conducted two-sample 
$Z$-tests to compare treatment–mediator and mediator–outcome effects across two groups.
Figure~\ref{fig:moti-ex} presents the estimated effects for a representative mediator (cg12682382), with heterogeneity test 
$p$-values of 0.081 and 0.017 for the two effects, respectively.
Our method identifies cg12682382 as a significant mediator, whereas homogeneous models fail to detect this signal, underscoring the importance of modeling heterogeneous mediation effects.

	Existing studies on heterogeneous mediation mainly focus on cross-sectional data using subgroup-based models \citep{wang2021causal,xue2022heterogeneous,dyachenko2023your}. However, the above methods assume a homogeneous pattern among mediators within the same subgroup, which may be restrictive when the subgroup structure varies across different mediators. Moreover, these methods assume a subgroup structure and may fail when there is non-cluster heterogeneity \citep{chernozhukov2018inference}. Additionally, they do not consider the dynamic heterogeneity. This motivates us to develop more flexible methods to capture the latent heterogeneity. 
	
	The individual heterogeneity is also related to the literature on unmeasured confounding. Traditional mediation analysis typically assumes that there are no unmeasured confounders \citep{imai2010identification,zhang2016estimating,zhao2022identification,guo2022high}. Several studies have contributed to addressing unmeasured confounders by incorporating latent factors \citep{guo2022doubly,yuanqu2023deconfounding}. For example, \cite{yuanqu2023deconfounding} removes the confounding bias by introducing a surrogate confounder based on multiple mediators. In the high-dimensional setting, \cite{guo2022doubly} develops a double-debiased procedure to correct both the confounding bias and regularization bias. However, these studies focus on cross-sectional settings and still assume homogeneous mediation effects.
	
	The second challenge arises from \textit{non-sparse mediators}. Mediation mechanisms can be highly complex and involve numerous mediators. Specifically, there may be a collective effect driven by the dense mediators. For example, in the context of DNA methylation, previous studies have demonstrated the presence of both site-specific hypermethylation and global hypomethylation \citep{reddington2014dna}. Consequently, the classical ``sparse-mediator assumption'' \citep{zhang2016estimating,perera2022hima2,guo2022high,zhao2022pathway} may not hold.

	To alleviate the constraints of the traditional sparsity assumption in high-dimensional data analysis, several factor-augmented variants have been developed, including the additive sparse and dense structure \citep{chernozhukov2017lava} and the conditional sparsity structure \citep{zheng2021nonsparse}. The main concept is that the coefficient vector becomes sparse once the effects of specific unobservable latent variables are adjusted. Nevertheless, these studies do not directly focus on mediation analysis and overlook interactions among high-dimension mediators.

	To tackle the aforementioned challenges, we propose an individualized dynamic mediation analysis (IDMA) method utilizing latent factors. Specifically, low-rank structures are imposed on the coefficients of linear structural equation models, enabling the disentanglement of individualized dynamic mediation effects into individual-specific dynamic factors and mediator-specific loadings. To capture the latent dynamics and borrow information across longitudinal measurements, we impose a fusion penalty on the individualized dynamic factors. A one-step convex estimation is developed based on initial estimates obtained from fixed-point iteration. 

    Recent works \citep{chakrabortty2018inference,luo2025multivariate} investigated multivariate mediation analysis by  modeling causal dependencies among mediators via directed acyclic graphs (DAGs), with the latter extending to longitudinal settings via time-varying and lagged structures. Our framework differs from theirs by focusing on heterogeneous mediation effects and considering non-causally correlated mediators arising from shared latent factors rather than direct causal links. 
    
	Our contribution to the mediation literature are threefold. First, by allowing the mediation effects to vary with individualized latent factors, our method adaptively captures both dynamic and individual-specific heterogeneity from data, generalizing subgroup-based approaches and accommodating distinct mediator-specific patterns. Second, we relax the traditional sparse-mediator assumption by introducing an sparse–dense interaction that captures collective mediation effects from dense mediators which might not be individually significant. Lastly, we establish the estimation consistency, mediator selection consistency, and asymptotic normality of the proposed estimator, providing new theoretical guarantees for heterogeneous and dynamic mediation analysis. Our theoretical results demonstrate the advantages of employing latent factors and leveraging temporal information.

	The remainder of the paper is structured as follows. Section \ref{sec:IDMA} outlines our IDMA methodology and its implementation is provided in \ref{sec:est}. Theoretical results are presented in Section \ref{sec:theory}, followed by numerical studies in Section \ref{sec:simu}. Section \ref{sec:appl} details the application of our method to the ADNI methylation data. Finally, we conclude our paper in Section \ref{sec:disc}.

	\section{Methodology}\label{sec:IDMA}
	We first introduce some notations. For any two real numbers $a$ and $b$, we write $a \vee b=\max (a, b)$. For a vector $\boldsymbol{\alpha}\in \mathbb{R}^{p}$, we use $\alpha_k\in \mathbb{R}$ to denote the $k$-th component of $\boldsymbol{\alpha}$. For vectors ${\boldsymbol{\alpha}}_{it}\in\mathbb{R}^{d}$ for ($i=1,\ldots,n$, $t=1,\ldots,T$), let $\bar {\boldsymbol{\alpha}}_i=({\boldsymbol{\alpha}}_{i1},\ldots,{\boldsymbol{\alpha}}_{iT})^{\top} \in\mathbb{R}^{T\times d}$ and $\{{\boldsymbol{\alpha}}_{it}\}_{1:n,1:T}=(\bar{\boldsymbol{\alpha}}_{1}^{\top},\ldots,\bar{\boldsymbol{\alpha}}_{n}^{\top})^{\top}\in\mathbb{R}^{nT\times d}$. For a matrix $\boldsymbol{B}$, let $\|\boldsymbol{B}\|_{2,0}$ denote the number of non-zero rows. Let $\lambda_{\max}(\boldsymbol{B})$ be the maximum eigenvalue of $\boldsymbol{B}$, and $\psi_{\max}(\boldsymbol{B})$ and $\psi_{\min}(\boldsymbol{B})$ be the maximum and minimum singular values of $\boldsymbol{B}$. We also use $\psi_j(\boldsymbol{B})$ to denote the $j$th largest singular value of $\boldsymbol{B}$. Moreover, we denote the Frobenius, operator, and nuclear norms of $\boldsymbol{B}$ as $\|\boldsymbol{B}\|_{F}$, $\|\boldsymbol{B}\|$, and $\|\boldsymbol{B}\|_{*}$ respectively. Let $[T]$ denote the index set $\{1,\ldots,T\}$ for any $T\in\mathbb{N}^{+}$. For any two matrices $\boldsymbol{A}$ and $\boldsymbol{B}$, we use $\boldsymbol{A} \otimes \boldsymbol{B}$ and $\boldsymbol{A} \odot \boldsymbol{B}$ to denote their Kronecker product and  Hardmard product, respectively, when they are of the same shape.
	
	\subsection{Motivation and Setup}
	In mediation analysis, we are interested in the effect of an exposure $X\in\mathbb{R}$ on the outcome variable $Y\in\mathbb{R}$ as well as its indirect effects through high-dimensional mediators ${\boldsymbol{M}}\in\mathbb{R}^p$, adjusted by the observed confounders ${\boldsymbol{Z}}\in\mathbb{R}^{q}$. In the literature on heterogeneous mediation analysis, the most commonly used approach is based on subgroup analysis \citep{xue2022heterogeneous}. Specifically, consider the following subgroup structural equation model
	\begin{equation}\label{eq:subsemY}
		Y_i = \theta_{h_i} X + \boldsymbol{\beta}^{\top}_{h_i} \boldsymbol{M}_i + \boldsymbol{\gamma}^{\top} Z_i + \varepsilon_i
	\end{equation}
	\begin{equation}\label{eq:subsemM}
		\boldsymbol{M}_i = \boldsymbol{\alpha}_{h_i} X + \boldsymbol{\Gamma} Z_i + \boldsymbol{\delta}_i
	\end{equation}
	for $i=1,\ldots,n$, where $h_i\in [H]$ represents the subgroup membership, and $H\in \mathbb{N}$ is the number of subgroups. Here $\theta_{h_i}\in\mathbb{R}$ reflects the group-specific direct effect of $X$ on $Y$,  ${\boldsymbol{\alpha}}_{h_i},{\boldsymbol{\beta}}_{h_i}\in\mathbb{R}^p$ are the group-specific mediation effects, and $\varepsilon_i$, $\boldsymbol{\delta}_i$ are random noises. The product ${\boldsymbol{\alpha}}_{h_i}^{\top}{\boldsymbol{\beta}}_{h_i}$ quantifies the indirect effect of $X$ on $Y$ within subgroup $h_i$. 
	
	The subgroup modeling allows mediation effects to vary across subgroups. Furthermore, the above subgroup structures are assumed to be shared by all mediation pathways. However, in practice, heterogeneity among individuals may not necessarily align with a fixed and homogeneous clustering structure, since the subgroup structures may vary across different mediation pathways and over time. This observation motivates us to develop new methods to better characterize heterogeneity for dynamic mediation analysis.

	To motivate our method, we can rewrite the coefficients in (\ref{eq:subsemY})-(\ref{eq:subsemM}) by introducing a subgroup indicator vector $\boldsymbol{e}_{h_i}\in\mathbb{R}^{H}$, where its $h_i$-th element is 1, and all other elements are 0. Let $\boldsymbol{a}_k=(\alpha_{1k},\ldots,\alpha_{Hk})^{\top}$, $\boldsymbol{b}_k=(\beta_{1k},\ldots,\beta_{Hk})^{\top}$, and $\boldsymbol{c}=(\theta_{1},\ldots,\theta_H)^{\top}$ denote the coefficient vector of the $k$-th mediator or the exposure on the $H$ subgroups. Then we have 
	\begin{equation}\label{eq:subgroup}
		\boldsymbol{\alpha}_{h_i}=\boldsymbol{A}\boldsymbol{e}_{h_i}, \quad \boldsymbol{\beta}_{h_i}=\boldsymbol{B}\boldsymbol{e}_{h_i}, \text{ and } \theta_{h_i}=\boldsymbol{c}^{\top}\boldsymbol{e}_{h_i}, \text{ for } i=1,\ldots,n.
	\end{equation}
	where $\boldsymbol{A}=(\boldsymbol{a}_1,\ldots,\boldsymbol{a}_p)^{\top}\in\mathbb{R}^{p\times H}$, and $\boldsymbol{B}=(\boldsymbol{b}_1,\ldots,\boldsymbol{b}_p)^{\top}\in\mathbb{R}^{p\times H}$. 
	This new formulation sheds light on the connection between subgroup-based mediation analysis and the existing literature on addressing unmeasured confounders \citep{wang2019blessings,guo2022doubly,yuanqu2023deconfounding,zhou2024promises}, as the subgroup indicator vector $\boldsymbol{e}_{h_i}$ functions analogously to a latent factor in these methods. Specifically, the above methods introduce a continuous latent factor $\boldsymbol{f}_i$ to capture hidden confounding in the $i$-th individual. The key idea is that when multiple parallel variables are conditional independent given the shared unobserved confounder $\boldsymbol{f}_i$, observing more parallel variables can facilitate the recovery of the underlying heterogeneity induced by $\{\boldsymbol{f}_i\}_{i=1}^{n}$. 
	
	In the context of mediation analysis, we are able to utilize multiple mediators to infer latent heterogeneity. Therefore, the latent factor model offers us a new strategy to identify latent heterogeneity in parameters and generalizes existing subgroup-based methods. Unlike the binary indicator vector $\boldsymbol{e}_{h_i}$ in (\ref{eq:subgroup}) that characterizes hard membership, continuous latent factors provide a more flexible approach, making it suitable when no explicit clustering structure exists for the mediation effects. Moreover, the continuous modeling approach enables a more straightforward way to incorporate dynamics.
	
	\subsection{Individualized Dynamic Mediation Analysis}\label{subsec:IDMA}
	We now focus on dynamic mediation analysis. For each individual $i$, $(Y_{it},{\boldsymbol{M}}_{it},X_{it},{\boldsymbol{Z}}_{it})$ are measured at time point $t$ for $t=1,\ldots,T$. 
	To capture the dynamic individual heterogeneity, we introduce a dynamic latent factor ${\boldsymbol{f}}_{it}\in \mathbb{R}^{r}$ for the $i$-th individual at time $t$. Specifically, we assume that potential outcomes at time $t$ depend solely on the treatment trajectory and/or the mediator trajectory at time $t$ after adjusting the observed confounder $\boldsymbol{Z}_{t}$ and the dynamic latent factors $\boldsymbol{f}_{t}$. Therefore, we denote the potential outcomes for $Y_{t}$ with exposure $X_t=x_t$ and mediator
	${\boldsymbol{M}}_t ={\boldsymbol{m}}_t$ by $Y_{t}(x_t ,{\boldsymbol{m}}_t)$. Similarly, denote the potential outcomes of ${\boldsymbol{M}}_{t}$ under exposure $X_t=x_t$ by ${\boldsymbol{M}}_{t}(x_t)$. 

    We introduce the following definitions on individualized treatment effects \citep{shalit2017estimating}, extending the averaged treatment effects in classical mediation analysis \citep{imai2010identification,imai2013identification} to the heterogeneous case. Defined the Individual Direct Effect (IDE) as 
$\operatorname{IDE}_{it}(x,x^{\prime})=\mathbb{E}[Y_{it}(x,\boldsymbol{M}_{it}({x^{\prime}}))-Y_{it}(x^{\prime},\boldsymbol{M}_{it}({x^{\prime}}))|\boldsymbol{f}_{it}]$, the Individual Indirect Effect (IIE) as 
$\operatorname{IIE}_{it}(x,x^{\prime})=\mathbb{E}[Y_{it}(x,\boldsymbol{M}_{it}({x}))-Y_{it}(x,\boldsymbol{M}_{it}({x^{\prime}}))|\boldsymbol{f}_{it}]$, and the Individual Indirect Effect through the $k$-th mediator ($\text{IIE}_k$) as $\operatorname{IIE}_{it}^{(k)}(x,x^{\prime})=\mathbb{E}[Y_{it}(x,\boldsymbol{M}_{it}({x}))-Y_{it}(x,({M}_{itk}({x^{\prime}}),\boldsymbol{M}_{it,-k}(x)))|\boldsymbol{f}_{it}]$ and $\boldsymbol{M}_{it,-k}$ represents the subvector of mediators excluding the $k$-th component.

    Without loss of generalizability, we omit the observed confounders ${\boldsymbol{Z}}$, as the extension of the following framework to include ${\boldsymbol{Z}}$ is straightforward. Taking individual heterogeneity into consideration, we propose the heterogeneous linear structural equations:
	\begin{equation}\label{eq:Ymodel}
		Y_{it}=\theta_{it}X_{it}+{\boldsymbol{\beta}}_{it}^{\top}{\boldsymbol{M}}_{it}+\varepsilon_{it},
	\end{equation}
	\begin{equation}\label{eq:Mmodel}
		{\boldsymbol{M}}_{it}={\boldsymbol{\alpha}}_{it}X_{it}+{\boldsymbol{\delta}}_{it},
	\end{equation}
	for $i=1,\ldots,n$, $t=1,\ldots,T$, where 
	$\theta_{it}\in\mathbb{R}$ reflects the effect of $X_{it}$ on $Y_{it}$, ${\boldsymbol{\alpha}}_{it},{\boldsymbol{\beta}}_{it}\in\mathbb{R}^p$ are the individualized dynamic mediation coefficients, and $\varepsilon_{it},{\boldsymbol{\delta}}_{it}$ are random errors with 0 mean and bounded variances. Moreover, the individualized coefficients can be decomposed by 
	${\alpha}_{itk}={\boldsymbol{a}}_{k}^{\top}{\boldsymbol{f}}_{it}$, ${\beta}_{itk}={\boldsymbol{b}}_{k}^{\top}{\boldsymbol{f}}_{it}$, for $k=1,\ldots,p$, and $\theta_{it}={\boldsymbol{c}}^{\top}{\boldsymbol{f}}_{it}$, where $ {\boldsymbol{a}}_k, {\boldsymbol{b}}_k$, and $\boldsymbol{c} \in \mathbb{R}^{r}$ are the factor loadings for the $k$-th mediator and the exposure $X$ respectively.

    {We make the following identification assumption for the individualized mediation effects. To facilitate exposition, we present only the model-based identification assumptions in the main text. The nonparametric identification assumptions and their implications, as well as their correspondence to the model-based counterpart below, are detailed in Section A of the Supplementary Material.

    \begin{assumption}\label{assump:identification}
        (i) Assume that (\ref{eq:Ymodel})-(\ref{eq:Mmodel}) hold and $r=O(1)$. Moreover, $(\varepsilon_{it},{\boldsymbol{\delta}}_{it})$ have continuous distributions and satisfying
        \begin{equation}\label{eq:indep-error}
   (\epsilon_{it},\boldsymbol{\delta}_{it}  ) \perp X_{it} | \boldsymbol{f}_{it};\quad \epsilon_{it} \perp \boldsymbol{\delta}_{it}| \boldsymbol{f}_{it}.
\end{equation}
        (ii) The errors satisfy
        \begin{equation}\label{eq:indep-error-2}
    \delta_{itk} \perp \boldsymbol{\delta}_{it,-k} |\boldsymbol{f}_{it},\quad\text{ for } k=1,\ldots,p.
\end{equation}
    \end{assumption}
In practice, Assumption \ref{assump:identification} reflects that the latent factor captures intrinsic
individual characteristics, such as genetic predisposition, stress response profile, or lifestyle
pattern, which jointly influence mediator dynamics and outcome trajectories. This is particularly relevant in applications like DNA methylation studies and psychosocial research. The low-rank assumption $r=O(1)$ ensures that the latent factors can be recovered from the dependency among observed mediators. 

Under Assumption \ref{assump:identification}, we have the following identification results.

\begin{proposition}
   Under Assumption \ref{assump:identification}(i), the individualized mediation effects $\operatorname{IDE}$ and $\operatorname{IIE}$  are identifiable through $\operatorname{IDE}_{it}(x,x^{\prime})=\theta_{it}(x-x^{\prime}), \operatorname{IIE}_{it}(x,x^{\prime})=\boldsymbol{\alpha}_{it}^{\top}\boldsymbol{\beta}_{it}(x-x^{\prime})$.
Further assume that Assumption \ref{assump:identification}(ii) holds. Then the individualized mediator-specific effect via the $k$-th mediator is identifiable through $\operatorname{IIE}^{(k)}(x,x^{\prime})=\alpha_{itk}\beta_{itk}(x-x^{\prime})$.
\end{proposition}
    Therefore, $\theta_{it}$, $\boldsymbol{\alpha}_{it}^{\top}\boldsymbol{\beta}_{it}$, and $\alpha_{itk}\beta_{itk}$ capture the individualized direct effect, the individualized total indirect effect through all mediators, and the individualized indirect effect through the $k$-th mediator, respectively, for a unit increase in treatment. }

    Unlike the traditional linear structural equation models \citep{zhang2016estimating,zhou2020estimation,perera2022hima2} and also the longitudinal structural equations models \citep{cai2022estimation} where the mediation effects are invariant among individuals, the parameters $(\boldsymbol{\alpha}_{it},\boldsymbol{\beta}_{it}, \theta_{it})$ in (\ref{eq:Ymodel}) and (\ref{eq:Mmodel}) are both individual-specific and time-specific, which allows heterogeneous dynamic mediation effects across individuals. 
	
	Further denote ${\mathcal{A}}=\{{\boldsymbol{\alpha}}_{it}\}_{1:n,1:T}\in\mathbb{R}^{nT\times p}$, ${\mathcal{B}}=\{{\boldsymbol{\beta}}_{it}\}_{1:n,1:T}\in\mathbb{R}^{nT\times p}$, ${\mathcal{C}}=\{{{\theta}}_{it}\}_{1:n,1:T}\in\mathbb{R}^{nT}$, $\boldsymbol{A}=({\boldsymbol{a}}_1,\ldots,{\boldsymbol{a}}_p)^{\top}\in\mathbb{R}^{p\times r}$, $\boldsymbol{B}=({\boldsymbol{b}}_1,\ldots,{\boldsymbol{b}}_p)^{\top}\in\mathbb{R}^{p\times r}$, and ${\boldsymbol{F}}=\{{\boldsymbol{f}}_{it}\}_{1:n,1:T}\in\mathbb{R}^{nT\times r}$. Intuitively, we are positing a low-rank structure of the individualized mediation coefficients, wherein 
	${\mathcal{A}}={\boldsymbol{F}}{\boldsymbol{A}}^{\top}$, ${\mathcal{B}}= {\boldsymbol{F}}{\boldsymbol{B}}^{\top}$, and  ${\mathcal{C}}={\boldsymbol{F}}{\boldsymbol{c}}$. In particular, we have
	\begin{equation}\label{eq:lowrank}
		\boldsymbol{\alpha}_{it}=\boldsymbol{A}\boldsymbol{f}_{it}, \boldsymbol{\beta}_{it}=\boldsymbol{B}\boldsymbol{f}_{it}, \text{ and } \boldsymbol{\theta}_{it}=\boldsymbol{c}^{\top}\boldsymbol{f}_{it}, \text{ for } i=1,\ldots,n, t=1,\ldots,T.
	\end{equation}
	This low-rank decomposition disentangles the personalized coefficients into two distinct components: individualized dynamic heterogeneity, characterized by the latent dynamic factor $\boldsymbol{f}_{it}$, and the mediator-specific loadings, captured by ${\boldsymbol{a}}_k$ and $\boldsymbol{b}_k$. For example, in the research of AD, these latent dynamic factors $\boldsymbol{f}_{it}$ include diverse brain pathological lesions, differences in cognitive reserve, genetic background, and environmental exposures \citep{dong2017heterogeneity}, all of which can change over time. 
	
	The loadings $\boldsymbol{a}_k$ and $\boldsymbol{b}_k$ characterize the sensitivity of the $k$-th mediator to the latent factors. For example, in DNA methylation research, studies have shown that the sensitivity of CpG sites to environmental influences varies significantly across the genome \citep{stepanyan2023long}. Some CpG sites are highly responsive to environmental factors, leading to changes in their methylation status, while others remain relatively stable regardless of environmental conditions.  Therefore, incorporating loadings provides a more accurate representation of the heterogeneity across different mediation pathways.
	
	A higher concordance between the individual factor $\boldsymbol{f}_{it}$ and the mediator loadings $(\boldsymbol{a}_k,\boldsymbol{b}_k)$ indicates that the $k$th mediation pathway is more important for the $i$th individual. It is reasonable to assume that a mediator deemed significant at the population level should have a non-degenerated effect across all individuals. Accordingly, selecting active mediators with nonzero individualized mediation effects is equivalent to identifying those with nonzero loadings $(\boldsymbol{a}_k,\boldsymbol{b}_k)$.
    Given that latent factors can vary across different individuals, and the factor loadings of different mediators may also vary, our proposed model allows for heterogeneous mediation effects for different mediation pathways. 
	

    \begin{figure}[h]
		\centering\includegraphics[width=4.5in]{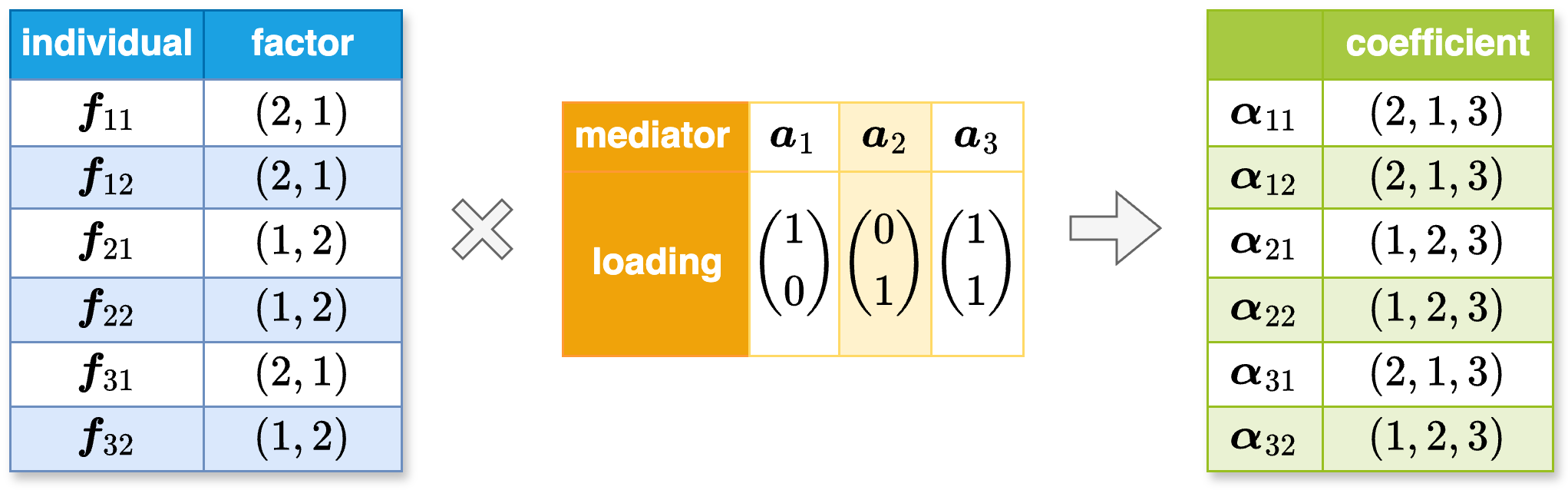}
		\caption{Heterogeneous mediation structure when $p=3, n=3, T=2$, and $r=2$.\label{fig:loading}}
	\end{figure}
	
	To illustrate, let's consider a simple case with $3$ mediators, $3$ individuals (each with $2$ measurements), and $2$-dimensional latent factors. Figure \ref{fig:loading} displays the coefficient loadings ${\boldsymbol{A}}\in\mathbb{R}^{p \times r}$ and the factor matrix $\boldsymbol{F}\in\mathbb{R}^{nT\times r}$. Note that the third mediator exhibits the same effect across all individuals, whereas the first two demonstrate heterogeneous effects. Additionally, we observe dynamic behavior for the third individual, whereas the first two individuals exhibit static characteristics. Hence, our model exhibits greater flexibility compared to existing methods assuming the same heterogeneous structure across all mediators.

	When a clear subgroup structure exists, we can 
	set $\boldsymbol{f}_{it} = \boldsymbol{e}_{h_{it}}$, where $h_{it}$ denotes the subgroup of individual $i$ at time $t$. In this case, (\ref{eq:lowrank}) reduces to (\ref{eq:subgroup}) in the subgroup-based mediation analysis. Therefore, our method can be viewed as a generalization of the subgroup-based mediation analysis to a more flexible soft clustering method, where heterogeneity is characterized by an $r$-dimensional continuous latent factor $\boldsymbol{f}_{it}$. As the number of mediators increases, we can recover the underlying latent factors ${\boldsymbol{f}}_{it}$ up to a rotation \citep{chernozhukov2018inference,chernozhukov2023inference}.

	\subsection{Mediation Selection with Latent Dynamics}\label{subsec:structure}
	The proposed disentanglement in Section \ref{subsec:IDMA} also provides additional advantages by enabling simultaneous mediator selection and dynamic factor modeling. 
	
	Specifically, the latent factor $\boldsymbol{f}_{it}$ captures the collective mediation effect from the dense mediators. Meanwhile, the site-specific mediation effects of mediators on the outcome are represented by their corresponding loadings $\boldsymbol{b}_k$. After adjusting the collective effect, we assume that the mediators with specific effects are sparse, i.e., 
	\begin{equation}\label{assump:sparse}
		\|\boldsymbol{B}\|_{2,0} \leq s_b.
	\end{equation} Unlike traditional high-dimensional mediation analyses which directly assume only a small subset of mediators are active \citep{zhao2022pathway,xue2022heterogeneous,shuai2023mediation}, our approach allows the majority of mediators to contribute to the outcome, with their collective effects summarized within the latent dynamic factors.
	
	Instead of modeling dynamics directly on the high-dimensional parameters $\boldsymbol{\alpha}$ and $\boldsymbol{\beta}$, we propose to characterize the dynamic mediation pattern via the latent factor $\boldsymbol{f}_{it}$. Specifically, denoting the dynamic factor matrix for the $i$th individual by $\bar{\boldsymbol{f}}_i=({\boldsymbol{f}}_{i1},\ldots,{\boldsymbol{f}}_{iT})^{\top}\in\mathbb{R}^{T\times r}$, we assume that
	\begin{equation}\label{assump:fused}
		\{\bar{\boldsymbol{f}}_i\}_{i=1}^{n}\in\left\{\bar{\boldsymbol{f}}_i \in\mathbb{R}^{T\times r} \left|\|D_f\operatorname{Vec}({\bar{\boldsymbol{f}}_i}^{\top})\|_0 \leq s_f\right.\right\},
	\end{equation}
	where $\operatorname{Vec}({\bar{\boldsymbol{f}}}^{\top}_i)=({\boldsymbol{f}}_{i1}^{\top},\ldots,{\boldsymbol{f}}_{iT}^{\top})^{\top}\in\mathbb{R}^{Tr}$ and
	\begin{equation}\label{eq:df}
		{D_f}=\left(\begin{array}{cccccc}
			-{\boldsymbol{I}}_r & {\boldsymbol{I}}_r &  &  & & \\
			& -{\boldsymbol{I}}_r & {\boldsymbol{I}}_r &  & & \\
			&  &  & \ldots & & \\
			&  &  &  & -{\boldsymbol{I}}_r & {\boldsymbol{I}}_r 
		\end{array}\right)\in\mathbb{R}^{(T-1)r \times Tr}.
	\end{equation}
	The adjacent homogeneity pursuit (\ref{assump:fused}) is imposed on the individualized dynamic latent factors, allowing us to leverage information across the temporal dimension. Note that when $T=2$ and $r=2$, $D_f\operatorname{Vec}({\bar{\boldsymbol{f}}_i}^{\top})={\boldsymbol{f}}_{i2}-{\boldsymbol{f}}_{i1}$. The  $L_0$ penalty term here indeed assumes similarity across adjacent time points. Here the first-order discrete difference operator $D_f$ can be substituted with higher-order difference operators to capture higher-order smoothness \citep{tibshirani2014adaptive}. 
	
	The fused penalty in (\ref{assump:fused}) implicitly employs a first-order latent Markov assumption to model the dynamics of mediation effects. Specifically, we assume that the dynamic changes observed in the mediation effects are fundamentally driven by the dynamic changes in latent factors. These latent factors follow a first-order Markov process \citep{bartolucci2009latent}, where the latent factor at time $t$ depends solely on the latent factor at time $t-1$. The latent Markov dynamic process has also been used in existing studies to characterize biological processes \citep{armond2014stochastic}.

    {\begin{remark}
Recent works \citep{chakrabortty2018inference,luo2025multivariate} model causal dependencies among mediators using DAGs and capture temporal dependence via first-order lagged structures. In contrast, our framework accounts for \textit{noncausal} dependencies induced by shared latent factors, which are common in high-dimensional biomedical studies where mediators are jointly influenced by unobserved genetic or environmental factors. Moreover, our method models temporal dynamics through a low-dimensional latent process $\boldsymbol{f}_{it}$ that evolves over time, avoiding the estimation and computational burden of high-dimensional lagged modeling. See Section~A of the Supplementary Material for a detailed comparison.
\end{remark}
}
	\section{Estimation}\label{sec:est}
	The core of our estimation procedure is a rotation argument. Since our target parameter of interest is the low-rank coefficient matrix, it suffices to recover the latent factors and loadings up to a rotation, as their product consistently estimates the coefficient matrix. 
	
	Denote ${\Theta}=(\mathcal{C},{\mathcal{B}}) \in \mathbb{R}^{nT\times (p+1)}$. Let ${\mathcal{L}}_{M}({{\mathcal{A}}})=\sum_{t=1}^{T}\left\|{\boldsymbol{M}}_{it}-{\boldsymbol{\alpha}}_{it}X_{it}\right\|_2^2$, and $
	{\mathcal{L}}_{Y}({{\Theta}})={\mathcal{L}}_{Y}(\mathcal{C},{\mathcal{B}})=\sum_{t=1}^{T}\left(Y_{it}-\theta_{it} X_{it} -{\boldsymbol{\beta}}_{it}^{\top}{\boldsymbol{M}}_{it}\right)^2$ denote the mediator and outcome loss functions, respectively. The estimation problem can then be formulated as
	\begin{equation}\label{opt:sparseIDMA}
		\begin{aligned}
			&\min_{{\Theta},{{\mathcal{A}}}}\quad {\mathcal{L}}_{Y}({\Theta})+{\mathcal{L}}_{M}({{\mathcal{A}}})
			\\\text{s.t. } & \text{rank}({{\mathcal{A}}})\leq r,\quad \text{rank}({\Theta})\leq r,
			\\& \|D_f\operatorname{Vec}({\bar{\boldsymbol{f}}}^{\top}_{i})\|_0 \leq s_f \quad (i=1,\ldots,n),\quad \|{\boldsymbol{B}}\|_{2,0} \leq s_b.
		\end{aligned}
	\end{equation}
	
	To make the estimation tractable, we propose to solve (\ref{opt:sparseIDMA}) in a three-step procedure:
	\begin{enumerate}[label= (\roman*)]
		\item obtain  $\hat{{\mathcal{A}}}$, $\hat{\boldsymbol{F}}$, and $\hat{\boldsymbol{A}}$ by solving
		\begin{equation}\label{eq:optMmodel1}
			\min_{{{\mathcal{A}}}} {\mathcal{L}}_{M}({{\mathcal{A}}}) \quad\text{ s.t. } \text{rank}({{\mathcal{A}}})\leq r,  \|D_f\operatorname{Vec}({\bar{\boldsymbol{f}}}^{\top}_{i})\|_0 \leq s_f \quad (i=1,\ldots,n).
		\end{equation}
		
		\item obtain $\hat{\boldsymbol{{\mathcal{B}}}}=\hat{\boldsymbol{F}}{\hat{\boldsymbol{{B}}}}^{\top}$ and $\hat{{\mathcal{C}}}=\hat{\boldsymbol{F}}\hat{{\boldsymbol{{c}}}}$ by solving 
		\begin{equation}\label{eq:optMmodel2}
			\min_{{\boldsymbol{B}},{\boldsymbol{c}}}{\mathcal{L}}_{Y}(\hat{\boldsymbol{F}}{\boldsymbol{{B}}}^{\top},\hat{\boldsymbol{F}}{\boldsymbol{{c}}}) \quad\text{ s.t. } \|{\boldsymbol{B}}\|_{2,0} \leq s_b.
		\end{equation}
        \item obtain the estimated individualized mediated effect through mediator $k$ by $\hat{\alpha}_{itk}\,\hat{\beta}_{itk}$ where $\hat{\alpha}_{itk}=\hat{\boldsymbol{f}}_{it}^{\top}\hat{\boldsymbol{a}}_k$, and $\hat{\beta}_{itk}=\hat{\boldsymbol{f}}_{it}^{\top}\hat{\boldsymbol{b}}_k$.

	\end{enumerate}
	
	Here in (\ref{eq:optMmodel1}) and (\ref{eq:optMmodel2}), the estimators $(\hat{\boldsymbol{{A}}},\hat{\boldsymbol{{B}}},\hat{\boldsymbol{{F}}})$ estimate the original parameter up to a rotation. However, the final estimators $\hat{{\mathcal{A}}}$ and $\hat{{\mathcal{B}}}$ are still consistent estimators of ${{\mathcal{A}}}$ and ${{\mathcal{B}}}$ respectively, as the rotation parts offset.

	The low-rank and $L_0$-sparse constraints in (\ref{eq:optMmodel1}) and (\ref{eq:optMmodel2}) lead to a NP-hard optimization problem. To incorporate low-rank constraint, we opt for nuclear norm penalization, the tightest convex relaxation of the NP-hard rank minimization problem \citep{candes2010matrix}. To incorporate the $L_0$-sparse assumptions on the latent factor and loadings, we apply SCAD penalty with local linear approximation \citep{zou2008one}. These relaxations ensure numerical stability and alleviate the nonconvexity of the original formulation. Further algorithmic details, optimization steps, and tuning procedures are provided in the Supplementary Material Section B.

	\section{Theoretical Properties}\label{sec:theory}
	In this section, we investigate the asymptotic properties of our IDMA estimator. Here we consider a more general case where $\beta_{itk}={\boldsymbol{b}}_k^{\top}{\boldsymbol{g}}_{it}$ and ${\boldsymbol{g}}_{it}$=${\boldsymbol{C}}{\boldsymbol{f}}_{it}$ for some invertible matrix ${\boldsymbol{C}} \in\mathbb{R}^{r\times r}$; that is, ${\boldsymbol{g}}_{it}$ is a rotation of ${\boldsymbol{f}}_{it}$. Therefore, $\mathcal{B}=\boldsymbol{G}\boldsymbol{B}^{\top}$ with ${\boldsymbol{G}}={\boldsymbol{F}}{\boldsymbol{C}}^{\top}$. Utilizing the singular value decomposition ${\mathcal{A}}={{U}}_A{{D}}_A{{V}}_A$ and ${\mathcal{B}}={{U}}_B{{D}}_B{{V}}_B$, we define $\boldsymbol{F}={U}_A{D}_A$, $\boldsymbol{A}={V}_A$, $\boldsymbol{G}={U}_B{D}_B$, and $\boldsymbol{B}={V}_B$. In the special case where $\boldsymbol{C}$ is an identity matrix, it aligns with the assumption outlined in Section  \ref{sec:IDMA}. In practice, the underlying mechanisms of factors between $\boldsymbol{M}$ and $Y$ and between $X$ and $\boldsymbol{M}$ may differ.
	Therefore, the rotation structure serves as a relaxation, enhancing the flexibility of our model. Since $\boldsymbol{C}$ is invertible, we have $\|{\boldsymbol{B}}\|_{2,0}=\|{\boldsymbol{B}}\boldsymbol{C}\|_{2,0}$, preserving the estimation procedure outlined in Section \ref{sec:est}.

	Let $\hat{\mathcal{A}}=\{\hat{\boldsymbol{\alpha}}_{it}\}_{1:n,1:T}$ and $\hat{\mathcal{B}}=\{\hat{\boldsymbol{\beta}}_{it}\}_{1:n,1:T}$ denote the estimator obtained by Algorithm 1 in Section B of the Supplementary Material. Assume that the parameter space for ${\bar{\boldsymbol{f}}_i}$ is 
	$$
	\mathcal{M}_{\boldsymbol{f}}^{(i)}=\left\{\bar{\boldsymbol{f}}_i\in\mathbb{R}^{T\times r}: {{f}}_{i,t-1,r^{\prime}}={{f}}_{i,t,r^{\prime}}, \text{ for } t\in \mathcal{T}^{(i)}_l, r^{\prime}=1,\ldots,r, l=1,\ldots,L\right\},
	$$
	where $\mathcal{T}^{(i)}_{l}\subseteq[T]$ denotes the index set of time points belonging to the $l$-th constant segment for individual $i$, and $\cup_{l=1}^{L}\mathcal{T}^{(i)}_{l}=[T]$. Without loss of generality, we assume that $\mathcal{T}^{(i)}_{l}$ does not depend on $r^{\prime}$ for notational simplicity. Let $\boldsymbol{\mu}_{i}\in\mathbb{R}^{L\times r}=[\boldsymbol{\mu}_{i1},\ldots,\boldsymbol{\mu}_{iL}]^{\top}$ denote the partitioned factor matrix, where the $l$th row represents a factor in the $l$th group. Let $Q_i\in\mathbb{R}^{T\times L}$ denote the indicator matrix, where $Q_i(t,l)=1$ represents $t\in\mathcal{T}^{(i)}_{l}$ for the $i$th individual and 0 otherwise. Then we have $\boldsymbol{f}_{i}=Q_i\boldsymbol{\mu}_{i}$  $(i=1,\ldots,n).$ Consider the oracle estimator of $\boldsymbol{\mu}_{i}$ when the partition $\{\mathcal{T}^{(i)}_{l}\}_{l=1}^{L}$ is known. 
    In fact, one can show that the adaptive estimators under unknown partitions will converge to the oracle estimators under regular assumptions. For notational simplicity, we do not consider the case here. Let $U_A=\left[{\boldsymbol{u}}_{11}, \ldots, {\boldsymbol{u}}_{nT}\right]^{\prime}$ and $V_A=\left[{\boldsymbol{v}}_1, \ldots, {\boldsymbol{v}}_p\right]^{\prime}$ denote the $nT \times r$ and $p \times$ $r$ matrices which represent the left singular vectors and right singular vectors of $\mathcal{A}$, respectively, corresponding to the nonzero singular values. We provide the following key assumptions.
	
	\begin{assumption}[Spiked singular value]\label{assump:ssv}
		The nonzero singular values of $\mathcal{A}$ are "spiked":
		$$
		\psi_1\left(\mathcal{A}\right) \geq \cdots \geq \psi_r\left(\mathcal{A}\right) \geq \psi_{n p}, \quad \psi_j\left(\mathcal{A}\right)=0, \text{ for any } j>r
		$$
		for some $\psi_{np}$ satisfying $((nT) \vee p)^{3/4}=o(\psi_{np})$. Assume that $\kappa=\psi_1(\mathcal{A})/\psi_r(\mathcal{A})=O(1).$
	\end{assumption}
	Assumption \ref{assump:ssv} here requires that the nonzero singular values are large enough to ensure the consistent estimation of the rank and accurate estimation of the singular vectors. This is necessary to establish entriwise convergence properties for the low-rank matrix. 
	
	\begin{assumption}[Incoherent singular vectors]\label{assump:isv}
		We assume incoherent singular-vectors:
		$$
		\mathbb{E}\max _{j \leq p}\left\|{\boldsymbol{v}}_j\right\|^2=O\left({r p^{-1}}\right), \quad \mathbb{E}\max_{i\leq n}\max_{t\leq T}\left\|{\boldsymbol{u}}_{it}\right\|^2=O\left({r {(nT)}^{-1}}\right).
		$$
	\end{assumption}
	Assumption \ref{assump:isv} prevents the information of the row and column
	spaces of the matrix from being too concentrated in a few rows or columns, which is known to be crucial for reliable
	recovery of low-rank matrices \citep{candes2010matrix,chen2020noisy}. 
	
	For a matrix $W$ with singular value decomposition $W=U_{W}D_{W}V_{W}$, let $P_{W}(\Delta)=U_{W} U_{W}^{\top} \Delta V_{W} V_{W}^{\top}$ and $P_{W}^{\perp}(\Delta)=\Delta-P_{W}(\Delta)$. Define the restricted set $\mathcal{R}^{W}_a=\{\Delta:\left\|P_{W}(\Delta)\right\|_{*}\leq a\left\|P_{W}^{\perp}(\Delta)\right\|_{*}\}$. Recall that $\tilde{\boldsymbol{X}}={\boldsymbol{X}}\otimes {\boldsymbol{1}}_{p}^{\top}$ where $\boldsymbol{X}\in\mathbb{R}^{nT}$ and  $\boldsymbol{X}_{T(i-1)+t}=X_{it}$.
	\begin{assumption}[Restricted Strong Convexity]\label{assump:rsc}
		For any $a>0$, there exists a constant $\kappa_a$, such that  $\|\tilde{\boldsymbol{X}}\odot \Delta\|_F^2\geq \kappa_a\|\Delta\|_F^2, \text{ for any } \Delta\in \mathcal{R}_a^{\mathcal{A}}$. 
	\end{assumption}
	
	The restricted strong convexity condition is commonly employed in regularized estimation. See \cite{negahban2012unified} for more details. The following theorem establishes the individual-wise convergence rates for the dynamic mediation coefficients $\boldsymbol{\alpha}_{it}$.

	\begin{theorem}\label{thm:1}
		Under Assumptions \ref{assump:ssv} - \ref{assump:rsc} and E.1-E.3 in Section E.1 of the Supplementary Material, if $\lambda_r \asymp \sqrt{nT+p}$, then for fixed $i\leq n$, fixed $k \leq p$, and fixed $t \leq T$, we have
		\begin{equation*}
			\left|\hat{\alpha}_{itk}-{\alpha}_{itk}\right|=O_p\left({{(nT)}}^{-1/2}+(p|\mathcal{T}_l^{(i)}|)^{-1/2}\right),
		\end{equation*}
        where $l$ satisfies that $t\in\mathcal{T}^{(i)}_l$.
	\end{theorem}

    
    {Let $\mathcal{S}:=\{k:\|\boldsymbol{a}_k\|_2\|\boldsymbol{b}_k\|_2>0\}$ denote the set of significant mediators with cardinality $s=|\mathcal{S}|$, and let $\hat{\mathcal{S}}:=\{k:\|\hat{\boldsymbol{a}}_k\|_2\|\hat{\boldsymbol{b}}_k\|_2>0\}$ be its estimation counterpart. We establish the selection consistency of $\hat{\mathcal{S}}$, as well as the joint asymptotic normality of $(\hat{\alpha}_{itk},\hat{\beta}_{itk})$ for $k\in\mathcal{S}$, together with their asymptotic covariance matrix, as stated in the following theorem. 
	
	\begin{theorem}\label{thm:2}
		Under Assumptions \ref{assump:ssv} - \ref{assump:rsc} and E.1-E.3 in Section E.1 of the Supplementary Material, if $\lambda_r \asymp \sqrt{nT+p}$ , $\lambda_b \rightarrow 0$, and $\sqrt{nT}\lambda_b\rightarrow \infty$, there exists a local minimizer $(\hat{\Theta},\hat{\mathcal{A}})$ of (\ref{opt:sparseIDMA}) that satisfies:

\noindent(1) $P(\hat{\alpha}_{itk}\hat{\beta}_{itk}=0)\rightarrow 1$ for $k \notin \mathcal{S}$, for each $i\in\{1,\ldots,n\}$, $t\in\{1,\ldots,T\}$, and

\noindent(2)
$\Sigma_{\alpha\beta}^{-1/2}
\begin{pmatrix}
\hat\alpha_{itk} - \alpha_{itk} \\
\hat\beta_{itk} - \beta_{itk}
\end{pmatrix}
\xrightarrow{d} 
N \!\left(
\boldsymbol{0},\mathbf{I}_2 
\right), \text{ with } \Sigma_{\alpha\beta}=\begin{pmatrix}
\sigma_{\alpha}^{2} & \rho_{\alpha\beta}\\[2pt]
\rho_{\alpha\beta} & \sigma_{\beta}^{2}
\end{pmatrix},
$
for $i\in \{1,\ldots,n\}$, $t\in \{1,\ldots,T\}$, and $k\in \mathcal{S}$, where the explicit form of $(\sigma_{\alpha}^{2},\sigma_{\beta}^{2},\rho_{\alpha\beta})$ is given in Section B.2.1 of the Supplementary Material.
	\end{theorem}
	 By applying the delta method, we can obtain the asymptotic property of the estimated individualized mediation effect $\hat{\alpha}_{itk}\hat{\beta}_{itk}$, as stated in the following corollary.

    \begin{corollary}\label{cor:asymvar}
        Under the assumptions in Theorem \ref{thm:2}, $\hat{\alpha}_{itk}\hat{\beta}_{itk}$ satisfies
\begin{equation}\label{eq:ab-asymp}    \left(\hat\alpha_{itk}\hat\beta_{itk} - \alpha_{itk}\beta_{itk} \right)/\sigma_{\alpha\beta}
\xrightarrow{d} 
N\!\left(0, 1\right),
\end{equation}
where $l$ satisfying that $t\in\mathcal{T}_l^{(i)}$ and
$\sigma^2_{\alpha\beta} 
= (\beta_{itj})^2 \sigma^2_{\alpha} 
+ (\alpha_{itj})^2 \sigma^2_{\beta} 
+ 2\alpha_{itj}\beta_{itj}\,\rho_{\alpha\beta}.$
    \end{corollary}
Based on Corollary \ref{cor:asymvar}, we can construct the $(1-\eta)\times 100\%$ confidence interval for the individual mediation effect ${\alpha}_{itk}{\beta}_{itk}$ as:
\begin{equation}\label{eq:CI}
    [\hat{\alpha}_{itk}\hat{{\beta}}_{itk} - z_{1-\eta/2}\hat{\sigma}_{\alpha\beta},
    \ \hat{\alpha}_{itk}\hat{{\beta}}_{itk} + z_{1-\eta/2}\hat{\sigma}_{\alpha\beta}],
\end{equation}
where $z_{1-\eta/2}$ denotes the $(1-\eta/2)$ quantile of the standard normal distribution and the estimate $\hat{\sigma}_{\alpha\beta}$ is provided in Section B.2 of the Supplementary Material.}
	
	\section{Simulation}\label{sec:simu}
	
	In the simulation, we evaluate three candidate mediation analysis methods. The first two are the methods proposed in \cite{zhang2016estimating} implemented with ``MCP'' (HIMA-MCP) and ``SCAD'' (HIMA-SCAD) penalties respectively. Additionally, we consider the debiased-lasso method (HIMA2) proposed in \cite{perera2022hima2}. We also compare the performance of the subgroup-based mediation method (gHIMA) proposed by \cite{xue2022heterogeneous}. Since these methods are not designed for longitudinal data, we fit each model using the data at each time point separately. 
	
	For each method, we use $\sqrt{\|\hat{\mathcal{A}}\odot\hat{\mathcal{B}}-{\mathcal{A}}\odot{\mathcal{B}}\|_F^2/(nTp)}$ to evaluate the root mean square error (RMSE). To evaluate the dynamic patterns captured by each estimator, we first define a soft trend operator $\mathbf{T}$ for a sequence $\boldsymbol{h}\in\mathbb{R}^{T}$ as $\mathbf{T}({\boldsymbol{h}})=\boldsymbol{d}\in\mathbb{R}^{T-1},$ where $d_{j}={I}(h_{j+1}-h_{j}>c) + (-1){I}(h_{j+1}-h_{j}< -c)$ $(j=1,\ldots,T-1)$ and some threshold $c$. The operator $\boldsymbol{T}$ transforms a continuous sequence to its corresponding trend vector. Here we choose $c$ as $0.1$ instead of $0$ to allow for slight disturbances over time. Let $\hat{\boldsymbol{\gamma}}_{i\cdot k}$ =$(\hat{{\alpha}}_{i1k}\hat{{\beta}}_{i1k},\ldots,\hat{{\alpha}}_{iTk}\hat{{\beta}}_{iTk})^{\top}$ be the coefficient estimators for the $i$th individual on the $k$th mediator. Define the true parameters ${\boldsymbol{\gamma}}_{i\cdot k}$ in the same way. Then we define the matching error (ME) for each estimator as 
	$
	\operatorname{ME}=\sqrt{{\sum_{i=1}^{n}\sum_{k=1}^{p}\|\mathbf{T}(\hat{\boldsymbol{\gamma}}_{i\cdot k})-\mathbf{T}({\boldsymbol{\gamma}}_{i\cdot k})\|^2}/{(np(T-1)})}
	$. Let ${\gamma}_{itk}={\alpha}_{i t k} {\beta}_{i t k}$ and $\hat{\gamma}_{itk}=\hat{\alpha}_{i t k} \hat{\beta}_{i t k}$. To evaluate selection accuracy on the individual level, we use the false negative ratio and false positive ratio over all parameters as follows
	\begin{footnotesize}
		$$\mathrm{FN.all}=\frac{\sum_{i=1}^n\sum_{t=1}^T\sum_{k=1}^p I\left(\hat{\gamma}_{itk}=0, {\gamma}_{itk} \neq 0\right)}{\sum_{i=1}^n\sum_{t=1}^T\sum_{k=1}^p I\left({\gamma}_{itk}  \neq 0\right)}, \quad \mathrm{FP.all}= \frac{\sum_{i=1}^n\sum_{t=1}^T\sum_{k=1}^p I\left(\hat{\gamma}_{itk} \neq 0, {\gamma}_{itk} = 0\right)}{\sum_{i=1}^n\sum_{t=1}^T\sum_{k=1}^p I\left({\gamma}_{itk} = 0\right)}.$$
	\end{footnotesize}
	Let $w_k=\sum_{i=1}^n\sum_{t=1}^T|{\gamma}_{itk} |$ and $\hat{w}_k=\sum_{i=1}^n\sum_{t=1}^T|\hat{\gamma}_{itk} |$. We  evaluate the selection accuracy on the population level by the averaged false negative ratio and false positive ratio
	\begin{footnotesize}
		$$
		\mathrm{FN.avg}=\frac{\sum_{k=1}^p I\left(\hat{w}_k=0, w_k  \neq 0\right)}{\sum_{k=1}^p I\left(w_k\neq 0\right)}, \quad \mathrm{FP.avg}= \frac{\sum_{k=1}^p I\left(\hat{w}_k\neq 0, w_k= 0\right)}{\sum_{k=1}^p I\left(w_k= 0\right)}.
		$$
	\end{footnotesize}
	
	We explore three distinct settings. Setting 1 represents a heterogeneous case where positive and negative individual effects coexist and offset each other at the population level, highlighting heterogeneity in the mediation process relevant to precision medicine \citep{jellinger2022recent}. Additionally, we assess the adaptivity of our method in a homogeneous setting (Setting 2), where all individuals share the same dynamic mediation effects. In Section C.1 of the Supplementary Material, we also examine another heterogeneous setting with no counteractive effects (Setting 3). In all settings, we set the number of time points as $T=5$ and generate the exposure $X_{it}$ from $\operatorname{Unif}[1,2]$. We consider $n=(50,100)$ and $p=(30,100)$. 

    Due to space limitations, additional simulations assessing robustness to nonlinear models, extra unmeasured confounders, varying $n/p$ ratios, and inference performance are reported in Section~C of the Supplementary Material.    
	
	\subsection{Heterogeneous setting with counteractive effects}\label{subsec:setting1}
	We first investigate the performance of all methods in a heterogeneous setting, where there are individually significant but collectively non-significant effects at the population level. Here we set $r=2$ and $\boldsymbol{c}=(1,0)^{\top}$.  Let $\bar{\boldsymbol{f}}_i^{(r^{\prime})}=(f_{i1r^{\prime}},\ldots,f_{iTr^{\prime}})^{\top}$ denote the $r^{\prime}$th column of $\bar{\boldsymbol{f}}_i$. For $i=1,\ldots,n$, we first generate ${\mu}_{i1},{\mu}_{i2}$ from $\mathcal{N}(1,0.15)$ independently and fix them. Then we set $\bar{\boldsymbol{f}}_i^{(1)}$ as $(\mu_{i1}\mathbf{1}_2,(\mu_{i1}-0.5)\mathbf{1}_3)^{\top}$ and $\bar{\boldsymbol{f}}_i^{(2)}$ as $w_i((\mu_{i2}-0.5)\mathbf{1}_2,\mu_{i2}\mathbf{1}_3)^{\top},$ where $w_i$'s are independent Radamacher variables.  We set ${\boldsymbol{b}}_{k}=(1,0)^{\top}$ for $k=1,3,5,7$; ${\boldsymbol{b}}_{k}=(0,1)^{\top}$ for $k=2,4,6,8$; and ${\boldsymbol{b}}_{k}=(0,0)^{\top}$ for $k=9,\ldots,p$. We also set the proportion of mediators with counteractive effects in the mediator model as $p_{\operatorname{hete}}=30\%$. For $k=1,\ldots,p$, we first generate $w_k\sim \operatorname{Bern}(1,p_{\operatorname{hete}}), h_k\sim \operatorname{Bern}(1,0.5),\mu_k\sim h_k\mathcal{N}(1.5,0.5)+(1-h_k)\mathcal{N}(-1.5,0.5)$, and then set ${\boldsymbol{a}}_{k}=(1-w_k)(\mu_k,0)^{\top}+w_k(0,\mu_k)^{\top}$. Since $\alpha_{itk}={\boldsymbol{a}}_k^{\top}{\boldsymbol{f}}_{it}$,  $\beta_{itk}={\boldsymbol{b}}_k^{\top}{\boldsymbol{f}}_{it}$, and $E[\bar{\boldsymbol{f}}_i^{(2)}]=0$, mediators with loadings ${\boldsymbol{a}}_k=(0,\mu_k)^{\top}$ (corresponding to $w_k=1$) and ${\boldsymbol{b}}_k=(0,1)^{\top}$ have counteractive effects on the population level. Therefore, it is difficult to detect them using a homogeneous model.

	The random error vectors $\operatorname{Vec}(\{\boldsymbol{\delta}_{it}\}_{t=1}^{T})$ are generated from two components ${\boldsymbol{\xi}}^{(1)}_{i}\in{\mathbb{R}}^{p} \sim \mathcal{N}({\boldsymbol{0}}_{p},\Sigma^M)$ and ${\boldsymbol{\xi}}_i^{(2)}\in{\mathbb{R}}^{T} \sim \mathcal{N}({\boldsymbol{0}}_{T},\Sigma^T)$, where $\Sigma^M_{k,k^{\prime}}={0.3}^{|k-k^{\prime}|}$ and $\Sigma^T_{t,t^{\prime}}=0.7\times{0.2}^{|t-t^{\prime}|}$. We let $\operatorname{Vec}(\{\boldsymbol{\delta}_{it}\}_{t=1}^{T})={\mathbf{1}}_{T}\otimes {\boldsymbol{\xi}}_i^{(1)}+{\boldsymbol{\xi}}_i^{(2)}\otimes {\mathbf{1}}_{p}$. We also generate $\varepsilon_{it}$ from $\mathcal{N}({\boldsymbol{0}}_{T},\Sigma^T)$. The simulation results are presented in Table \ref{tab:simu-hete1}.
	In general, the IDMA outperforms all the other candidates under all measures across different $n$ and $p$. We highlight four points here.

	First, the IDMA reduces the RMSE of those homogeneous methods by over 65\%, with significantly improved false negative and false positive ratios of both the population and individual levels. It effectively controls both FP.avg and FP.all below level 0.03 when $n=50$ and below level 0.005 when $n=100$, respectively, while maintaining much smaller FN.avg and FN.all. This indicates that the IDMA excels at detecting heterogeneity among individuals and selecting mediators in a personalized manner. In contrast, homogeneous methods only identify significant mediators at a population level, resulting in a high false negative ratio. This is due to the presence of $30\%$ of mediators whose effects offset at the population level despite their significance at the individual level. Detecting these mediators with a homogeneous model proves challenging, even with a large sample size.  In contrast, the IDMA retrieves individualized effects through low-rank decomposition, enabling the detection of significant mediators even if their effects offset at the population level.

	\begin{table}[H]
		\caption{Performance of the proposed method and competing methods in the heterogeneous setting with counteractive effects. \label{tab:simu-hete1}}
		\resizebox{\textwidth}{!}{\begin{tabular}{c|c|c|c|c|c|c|c|c}
				\hline\hline
				$p$                     & $n$                    & Method        & RMSE                  & ME                    & FN.avg                & FP.avg                & FN.all                & FP.all                \\ \hline
				\multirow{10}{*}{30}  & \multirow{5}{*}{50}  & \textbf{IDMA} & \textbf{0.261(0.077)} & \textbf{0.212(0.074)} & \textbf{0.03(0.067)}  & \textbf{0.026(0.035)} & \textbf{0.03(0.067)}  & \textbf{0.026(0.035)} \\ 
				&                      & HIMA-MCP      & 0.796(0.088)          & 0.513(0.062)          & 0.369(0.12)           & 0.429(0.131)          & 0.723(0.059)          & 0.116(0.04)           \\ 
				&                      & HIMA-SCAD     & 0.811(0.101)          & 0.539(0.072)          & 0.361(0.119)          & 0.515(0.137)          & 0.704(0.064)          & 0.147(0.05)           \\ 
				&                      & HIMA2         & 0.838(0.071)          & 0.901(0.034)          & 0.001(0.012)          & 0.999(0.006)          & 0.147(0.05)           & 0.872(0.018)          \\ 
				&                      & gHIMA         & 1.183(0.188)          & 0.831(0.054)          & 0(0)                  & 1(0)                  & 0.554(0.055)          & 0.441(0.056)          \\ \cline{2-9} 
				& \multirow{5}{*}{100} & \textbf{IDMA} & \textbf{0.194(0.03)}  & \textbf{0.178(0.058)} & \textbf{0.007(0.03)}  & \textbf{0.003(0.011)} & \textbf{0.007(0.03)}  & \textbf{0.003(0.011)} \\ 
				&                      & HIMA-MCP      & 0.697(0.038)          & 0.485(0.054)          & 0.406(0.086)          & 0.394(0.119)          & 0.689(0.053)          & 0.111(0.037)          \\  
				&                      & HIMA-SCAD     & 0.698(0.037)          & 0.517(0.06)           & 0.374(0.101)          & 0.512(0.129)          & 0.648(0.054)          & 0.155(0.049)          \\ 
				&                      & HIMA2         & 0.724(0.036)          & 0.871(0.032)          & 0(0)                  & 1(0)                  & 0(0)                  & 1(0)                  \\ 
				&                      & gHIMA         & 0.821(0.047)          & 0.703(0.047)          & 0(0)                  & 1(0)                  & 0.694(0.042)          & 0.297(0.042)          \\ \hline
				\multirow{10}{*}{100} & \multirow{5}{*}{50}  & \textbf{IDMA} & \textbf{0.106(0.07)}  & \textbf{0.072(0.05)}  & \textbf{0.021(0.062)} & \textbf{0.029(0.041)} & \textbf{0.021(0.062)} & \textbf{0.029(0.041)} \\ 
				&                      & HIMA-MCP      & 0.415(0.047)          & 0.257(0.031)          & 0.554(0.12)           & 0.104(0.038)          & 0.845(0.051)          & 0.024(0.009)          \\  
				&                      & HIMA-SCAD     & 0.416(0.047)          & 0.269(0.035)          & 0.52(0.115)           & 0.13(0.045)           & 0.83(0.055)           & 0.03(0.011)           \\ 
				&                      & HIMA2         & 0.522(0.044)          & 0.602(0.019)          & 0.402(0.123)          & 0.593(0.054)          & 0.656(0.06)           & 0.253(0.005)          \\  
				&                      & gHIMA         & 0.717(0.066)          & 0.6(0.026)            & 0(0)                  & 1(0)                  & 0.79(0.026)           & 0.206(0.02)           \\ \cline{2-9} 
				& \multirow{5}{*}{100} & \textbf{IDMA} & \textbf{0.061(0.009)} & \textbf{0.038(0.023)} & \textbf{0(0)}         & \textbf{0.005(0.009)} & \textbf{0(0)}         & \textbf{0.005(0.009)} \\ 
				&                      & HIMA-MCP      & 0.354(0.019)          & 0.24(0.029)           & 0.465(0.078)          & 0.085(0.031)          & 0.757(0.044)          & 0.019(0.006)          \\  
				&                      & HIMA-SCAD     & 0.351(0.019)          & 0.234(0.04)           & 0.449(0.082)          & 0.107(0.049)          & 0.747(0.054)          & 0.024(0.012)          \\  
				&                      & HIMA2         & 0.443(0.026)          & 0.687(0.018)          & 0.392(0.099)          & 0.694(0.049)          & 0.533(0.044)          & 0.438(0.004)          \\ 
				&                      & gHIMA         & 0.593(0.044)          & 0.599(0.033)          & 0(0)                  & 1(0)                  & 0.784(0.026)          & 0.215(0.025)          \\ \hline\hline
		\end{tabular}}
		
	\end{table}

	Second, the IDMA exhibits the smallest ME across all scenarios, indicating its superior ability to capture dynamic patterns of mediation effects. This can be attributed to two factors: the low-rank structure aids in recovering the underlying dynamic latent factors, facilitating the capture of dynamic changes in coefficients; and the fused penalty imposed on dynamic factors further enhances information borrowing from adjacent time points.
	
	Third, the subgroup-based method gHIMA performs even worse than the homogeneous method in the heterogeneous settings. This is due to the absence of a subgroup structure in the examined heterogeneous setting. In contrast, the IDMA effectively captures heterogeneity from the low-rank structure, accommodating various heterogeneous mechanisms.
	
	Last, the performance of the IDMA significantly improves with increased sample size. Particularly in the high-dimensional setting when $ p = 100 $, both RMSE and ME of the IDMA are reduced by around 50\% when $n$ increases from 50 to 100. In contrast, the decreases in RMSE and ME in the other methods are less than 14\% and 8\%, respectively.

	\subsection{Homogeneous setting}\label{subsec:setting2}
	We consider a homogeneous setting to evaluate the robustness of the IDMA. Let $r=2$. We generate ${\boldsymbol{a}}_{k}$ from $w_k\times\mathcal{N}((1,0)^{\top},0.1\times{\mathbf{I}}_2)+(1-w_k)\times\mathcal{N}((0,1)^{\top},0.1\times{\mathbf{I}}_2)$ with $w_k\sim \operatorname{Bern}(1,0.5)$ for $k=1,\ldots,p$, and fix them. Let $\boldsymbol{c}=(1,0)^{\top}$. For $p=30$, we generate ${\boldsymbol{b}}_{k}$ in the same way as that for ${\boldsymbol{a}}_k$ and further set ${\boldsymbol{b}}_k=\boldsymbol{0}$ if $\|{\boldsymbol{b}}_k\|^2_2<0.8$. For $p=100$, we set ${\boldsymbol{b}}_{1}={\boldsymbol{b}}_{2}=(1.5,0)^{\top}$, ${\boldsymbol{b}}_{3}={\boldsymbol{b}}_{4}=(0,1.5)^{\top}$, and ${\boldsymbol{b}}_{k}=(0,0)^{\top}$ for $k=5,\ldots,p$.
	The latent factors for all $i=1,\ldots,n$ are set as ${\boldsymbol{f}}_{it}=(1.5,0.1)^{\top}$ for $t=1,2,3$
	and ${\boldsymbol{f}}_{it}=(0.1,1.5)^{\top}$ for $t=4,5$.
	The $\epsilon_{it}$'s are the same as those in Section \ref{subsec:setting1}, while the $\boldsymbol{\delta}_{it}$'s are generated with $\Sigma^M$ and $\Sigma^T$ replaced by $\Sigma^M_{k,k^{\prime}}=0.5\times{0.3}^{|k-k^{\prime}|}$ and $\Sigma^T_{t,t^{\prime}}=0.5\times{0.2}^{|t-t^{\prime}|}$, respectively. The results of all the candidate methods are presented in Table \ref{tab:simuhomo}.

	As evident from Table \ref{tab:simuhomo}, the IDMA demonstrates robustness in the homogeneous setting and exhibits the best performance among all candidates across different $ n $ and $ p $. Robustness is particularly crucial in practical scenarios, as we are often uncertain about the degree of heterogeneity among individuals. The notable improvement over existing methods is due to the utilization of the low-rank structure in estimating the mediator model (\ref{eq:Mmodel}). This enables us to leverage the correlation structure among mediators from the same individual to generate more accurate estimates of $\boldsymbol{\alpha}_{it}$. Furthermore, the fusion penalty allows for information borrowing from adjacent time points. Additionally, we observe that the performance of IDMA further enhances with increased sample size. Specifically, the RMSE of IDMA decreases by 25\% when the sample size is doubled from 50 to 100.
	
	\begin{table}[H]
		\caption{Performance of the proposed method and competing methods in the homogeneous setting.\label{tab:simuhomo}}
		\resizebox{\textwidth}{!}{
			\begin{tabular}{c|c|c|c|c|c|c|c|c}
				\hline\hline
				$p$                     & $n$                    & Method        & RMSE                  & ME                    & FN.avg        & FP.avg                & FN.all        & FP.all                \\ \hline
				\multirow{10}{*}{30}  & \multirow{5}{*}{50}  & \textbf{IDMA} & \textbf{0.126(0.026)} & \textbf{0.285(0.079)} & \textbf{0(0)} & \textbf{0(0)}         & \textbf{0(0)} & \textbf{0(0)}         \\  
				&                      & HIMA-MCP      & 0.649(0.112)          & 0.577(0.056)          & 0(0)          & 0.412(0.137)          & 0.3(0.068)    & 0.115(0.046)          \\  
				&                      & HIMA-SCAD     & 0.649(0.113)          & 0.574(0.065)          & 0(0)          & 0.443(0.166)          & 0.301(0.07)   & 0.128(0.06)           \\ 
				&                      & HIMA2         & 0.675(0.09)           & 0.855(0.042)          & 0(0)          & 1(0.005)              & 0.11(0.038)   & 0.858(0.014)          \\ 
				&                      & gHIMA         & 0.92(0.086)           & 0.535(0.047)          & 0(0)          & 1(0)                  & 0.835(0.036)  & 0.184(0.035)          \\ \cline{2-9} 
				& \multirow{5}{*}{100} & \textbf{IDMA} & \textbf{0.097(0.017)} & \textbf{0.262(0.074)} & \textbf{0(0)} & \textbf{0(0)}         & \textbf{0(0)} & \textbf{0(0)}         \\ 
				&                      & HIMA-MCP      & 0.363(0.067)          & 0.489(0.043)          & 0(0)          & 0.211(0.115)          & 0.14(0.027)   & 0.048(0.029)          \\ 
				&                      & HIMA-SCAD     & 0.363(0.068)          & 0.478(0.045)          & 0(0)          & 0.207(0.139)          & 0.148(0.03)   & 0.046(0.033)          \\ 
				&                      & HIMA2         & 0.371(0.065)          & 0.689(0.042)          & 0(0)          & 1(0)                  & 0(0)          & 1(0)                  \\ 
				&                      & gHIMA         & 0.84(0.049)           & 0.529(0.031)          & 0(0)          & 1(0)                  & 0.804(0.027)  & 0.191(0.027)          \\ \hline
				\multirow{10}{*}{100} & \multirow{5}{*}{50}  & \textbf{IDMA} & \textbf{0.066(0.013)} & \textbf{0.126(0.033)} & \textbf{0(0)} & \textbf{0.004(0.008)} & \textbf{0(0)} & \textbf{0.004(0.008)} \\ 
				&                      & HIMA-MCP      & 0.22(0.061)           & 0.244(0.036)          & 0(0)          & 0.107(0.04)           & 0.398(0.063)  & 0.024(0.01)           \\ 
				&                      & HIMA-SCAD     & 0.219(0.061)          & 0.228(0.04)           & 0(0)          & 0.104(0.054)          & 0.416(0.062)  & 0.023(0.012)          \\ 
				&                      & HIMA2         & 0.423(0.079)          & 0.582(0.019)          & 0.015(0.06)   & 0.679(0.035)          & 0.488(0.079)  & 0.25(0.003)           \\ 
				&                      & gHIMA         & 0.547(0.101)          & 0.148(0.038)          & 0.155(0.279)  & 0.714(0.423)          & 0.842(0.119)  & 0.008(0.007)          \\ \cline{2-9} 
				& \multirow{5}{*}{100} & \textbf{IDMA} & \textbf{0.049(0.012)} & \textbf{0.11(0.035)}  & \textbf{0(0)} & \textbf{0.002(0.005)} & \textbf{0(0)} & \textbf{0.002(0.005)} \\  
				&                      & HIMA-MCP      & 0.154(0.039)          & 0.219(0.036)          & 0(0)          & 0.093(0.049)          & 0.321(0.067)  & 0.02(0.011)           \\ 
				&                      & HIMA-SCAD     & 0.154(0.04)           & 0.185(0.033)          & 0(0)          & 0.06(0.045)           & 0.37(0.074)   & 0.013(0.01)           \\ 
				&                      & HIMA2         & 0.311(0.043)          & 0.654(0.028)          & 0(0)          & 0.908(0.024)          & 0.374(0.069)  & 0.432(0.003)          \\ 
				&                      & gHIMA         & 0.479(0.067)          & 0.128(0.029)          & 0.227(0.302)  & 0.62(0.488)           & 0.812(0.086)  & 0.004(0.005)          \\ \hline\hline
		\end{tabular}}
		
	\end{table}
	
	\section{Application to ADNI methylation data}
	\label{sec:appl}
In this section, we analyze the microarray whole-genome DNA methylation profiles of participants enrolled in  Alzheimer’s Disease Neuroimaging Initiative (ADNI, adni.loni.usc.edu).  Our goal is to estimate the dynamic effect of the geriatric depression scale (GDS, serving as $\boldsymbol{X}$), on the progress of Alzheimer’s Disease (AD) mediated by the DNA methylation levels of different CpG sites (serving as $\boldsymbol{M}$), and identify significant CpG sites for mediators. As DNA-m is a reversible process, targeting DNA-m with drugs holds promise for reversing disease progression  \citep{bhootra2023dna}. Therefore, identifying DNA methylations on specific CpG sites as mediators is essential for developing targeted therapies in disease management.
	
Recent studies have recognized Alzheimer's disease as a pathologically heterogeneous disorder with different forms of cognitive presentation \citep{avelar2023decoding}. This heterogeneity is shaped by various factors, including genetic, demographic, and neuropsychiatric influences \citep{koulouri2024epigenetics}. Therefore, a heterogeneous dynamic mechanism underlies how individual experiences impact DNA-m, which in turn influences dementia pathogenesis.
	
In assessing the degree of cognitive dysfunction in Alzheimer’s disease, we use the 13-item version of the Alzheimer’s Disease Assessment Scale–Cognitive Subscale (ADAS-Cog 13, serving as $Y$). This scale investigates assessments of attention and concentration, planning and executive function, verbal and nonverbal memory, praxis, and delayed word recall, as well as number cancellation or maze tasks. Participants receive scores ranging from 0 to 85, with higher scores indicating poorer cognitive performance.

	\subsection{Prediction results}\label{subsec:applpred}
	After preprocessing (see Section D.1 of the Supplementary Material for the details), there are 217 individuals in our dataset, each with 3 observations of their methylation levels at 112 CpG sites collected in years 2010, 2011, and 2012, respectively, and their outcomes are standardized. To evaluate the validity of our proposed method, we randomly choose 80\%
	of the individuals as the training dataset, and compare the prediction errors of IDMA with those of the candidates mentioned in Section \ref{sec:simu} on the remaining test dataset. The above procedure is repeated 100 times. 
	
	For IDMA, we set $r=1$ by conducting a principal component analysis on the mediator methylation matrix and examining the cumulative variance contribution of the leading components. Then we apply the proposed estimation procedure in Section \ref{sec:est} to the training dataset to obtain the estimated individualized dynamic factors $\{\hat{\boldsymbol{f}}_{it}\}_{i \in \mathcal{D}_{\text{train}}}$ and loading matrices $\hat{\boldsymbol{A}}$ and $\hat{\boldsymbol{B}}$. To enable out-of-sample prediction for our IDMA, we fit a random forest using the estimated latent factors from training individuals as responses and their corresponding observed mediators along with demographic covariates (age, gender, and education) as predictors. The trained model is then used to predict the dynamic latent factors $\{\hat{\boldsymbol{f}}_{it}\}_{i \in \mathcal{D}_{\text{test}}}$ for new individuals, which are combined with the $\hat{\boldsymbol{A}}$ and $\hat{\boldsymbol{B}}$ to obtain individualized coefficients $\hat{\boldsymbol{\alpha}}_{it}=\hat{\boldsymbol{A}}\hat{\boldsymbol{f}}_{it}$, and $\hat{\boldsymbol{\beta}}_{it}=\hat{\boldsymbol{B}}\hat{\boldsymbol{f}}_{it}$ and to predict mediators and outcomes in the test set. Table \ref{tab:applpred} presents the root mean square errors of the fitted mediator model and the outcome model (RMSE.M and RMSE.Y), the prediction errors of the mediator model and the outcome model on the test dataset (PMSE.M and PMSE.Y), and the number of mediators with non-zero coefficients (No.Med). 
	
	Table \ref{tab:applpred} shows that our IDMA demonstrates superior performance compared to all other candidates in terms of both RMSE and PMSE for both the mediator model and the outcome model. Notably, IDMA achieves reductions in RMSE and PMSE for the mediator model and the outcome model by approximately 25\% and 50\%, respectively, in contrast to homogeneous methods. The subgroup-based mediation analysis method (i.e., gHIMA) emerges as the second-best method among all candidates. This suggests the presence of heterogeneity among individuals, where homogeneous methods fall short. Additionally, heterogeneous methods yield a more parsimonious model, incorporating only 29 mediators, compared to the models obtained from homogeneous methods.

    \begin{table}[H]
		\centering
		\caption{Prediction errors of the proposed method and competing methods on the test datasets over 100 replications.\label{tab:applpred}}
		\resizebox{0.8\textwidth}{!}{\begin{tabular}{c|c|c|c|c|c}
				\hline\hline
				Method        & RMSE.M                & PMSE.M               & RMSE.Y                & PMSE.Y                & No.Med                \\ \hline
				\textbf{IDMA} & \textbf{1.538(0.059)} & \textbf{1.58(0.048)} & \textbf{0.693(0.055)} & \textbf{0.917(0.071)} & \textbf{29.08(4.634)} \\ \hline
				HIMA-MCP      & 2.063(0.005)          & 2.064(0.02)          & 1.742(0.575)          & 1.893(0.541)          & 49.065(3.876)         \\ \hline
				HIMA\_SCAD    & 2.063(0.005)          & 2.064(0.02)          & 1.641(0.564)          & 1.769(0.546)          & 52.08(10.119)         \\ \hline
				HIMA2         & 2.08(0.008)           & 2.08(0.02)           & 2.03(0.377)           & 2.051(0.39)           & 96.81(2.672)          \\ \hline
				gHIMA         & 2.065(0.006)          & 2.069(0.025)         & 1.632(0.487)          & 1.031(0.247)          & 29.24(4.626)          \\ \hline\hline
		\end{tabular}}
		
	\end{table}

	Furthermore, IDMA outperforms gHIMA in both in-sample fitness and prediction accuracy of the mediator model, demonstrating its superior ability to capture individual-level heterogeneity. Specifically, IDMA identifies heterogeneous patterns in 44 out of 100 replications, whereas gHIMA detects only limited subgroup structures across years. This underestimation of heterogeneity explains gHIMA’s weaker performance. The improved results of IDMA highlight the benefits of incorporating a low-rank structure and a temporal fused penalty, which together enhance information sharing across mediation pathways and time points to better reveal underlying heterogeneity.
    
	\subsection{{Confidence intervals for individualized mediation effects}}\label{subsec:applinf}
\begin{figure}[H]
    \centering
    \includegraphics[width=0.8\linewidth]{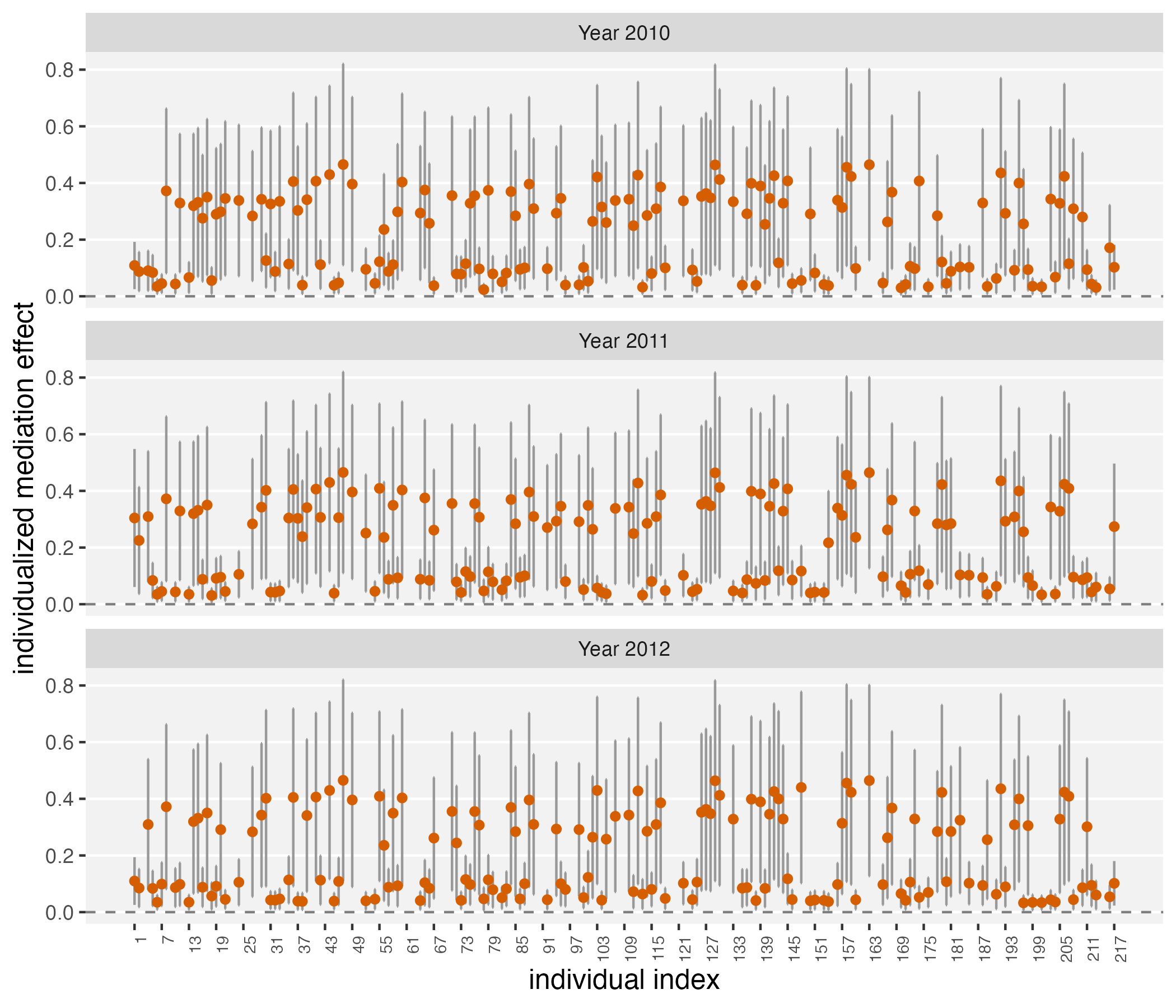}
    \caption{Estimated individualized mediation effects of GDS on the degree of cognitive dysfunction in AD via CpG site ``cg01343084'' in 2010, 2011, and 2012, with corresponding 95\% confidence intervals.}
    \label{fig:adni-ci}
\end{figure}

    {For the training samples with observed mediators and outcomes, we apply the inference procedure described in Section B.2.2 to estimate the variance of individualized mediation effects and construct corresponding confidence intervals. Figure~\ref{fig:adni-ci} presents the estimated effects of GDS on cognitive dysfunction in AD through a CPG site identified exclusively by our method. Information on genes identified by different methods is detailed in Section D.3 of the Supplementary Material.  As shown, all confidence intervals exclude zero, suggesting consistently significant and nonzero individualized mediation effects over time. }

	\subsection{Identified dynamic latent factors}\label{subsec:idlf}
\begin{figure}[h]
		\centering
		\includegraphics[width=0.9\textwidth]{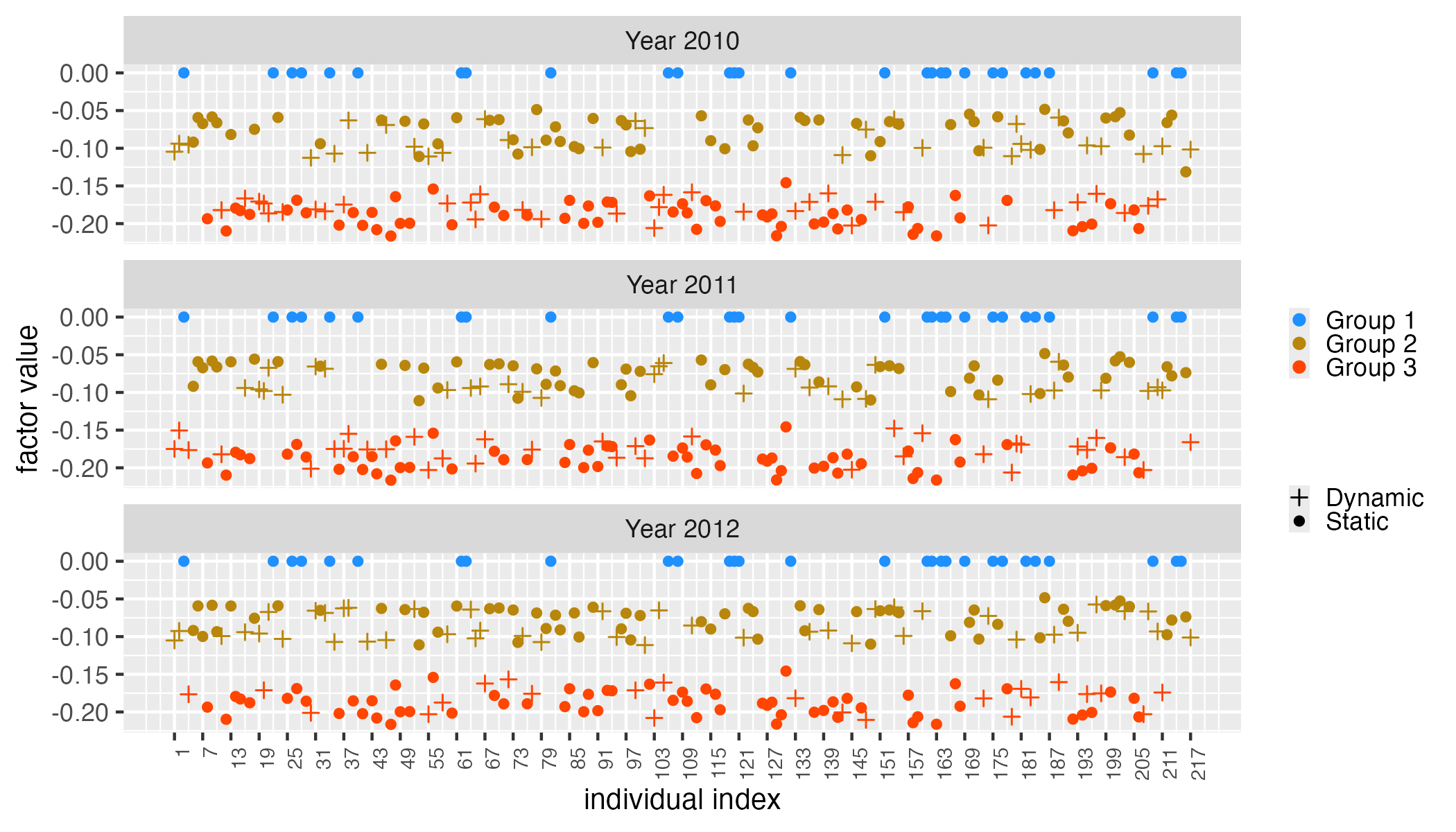}
		\caption{Recovered individual dynamic latent factor with group membership as well as membership patterns from 2010 to 2012. ``Dynamic'' indicates that the membership changes over time and ``Static'' indicates that the membership remains invariant over time.}
		\label{fig:dynamic_factor}
	\end{figure}

	We apply IDMA to the entire dataset to derive the estimated individualized dynamic latent factors $\{\{\hat{\boldsymbol{f}}_{it}\}_{t=1}^{3}\}_{i=1}^{217}$ for all participants. Subsequently, we employ K-means clustering at each time point and identify three subgroups. Out of the 217 individuals, 66 exhibit varying subgroup membership across different years, while the remaining 151 maintain static membership across the three levels. Figure \ref{fig:dynamic_factor} plots the individualized dynamic latent factors $\hat{\boldsymbol{f}}_{it}$ across all subjects $i\in \{1,\ldots,217\}$ and time points $t=1,2,3,$ corresponding to year 2010, 2011, 2012. Point color corresponds to the group membership of the corresponding individual; points marked with “$+$” indicate individuals whose membership changes across 2010–2012, while points marked with “$\cdot$” represent individuals whose membership does not change. More details about the identified groups are provided in Section D.2.

	Unlike gHIMA, which performs subgroup analysis directly on mediation effects, our method clusters individuals in the latent space obtained from the low-rank decomposition of the coefficient matrix. As noted in Section \ref{subsec:applpred}, gHIMA only identifies the 3-group structure in 2 out of 100 replications, suggesting that the subgroup structure might not be evident at the coefficient level but clearer in the recovered latent space.

	In general, the proposed IDMA demonstrates superior predictive performance in both the mediator and outcome models while maintaining a more parsimonious structure. Notably, our approach successfully identifies genes that have recently been reported as highly associated with AD, while also highlighting potential candidate genes related to AD. The individualized dynamic mediation effects of detected CpG sites are provided in Section D.4.
	\section{Discussion}\label{sec:disc}
	{We conclude the paper with two remarks. Firstly, in our framework, sparsity is imposed only on the outcome loading matrix $\boldsymbol{B}$ to enable mediator selection, while the treatment–mediator effects are estimated through a low-rank factorization without additional shrinkage. This design is justified since the treatment–mediator relation forms a well-posed low-dimensional problem under the low-rank assumption. Nevertheless, the framework can be extended by adding sparsity to both $\boldsymbol{A}$ and $\boldsymbol{B}$ or introducing a joint penalty on $\{\|\boldsymbol{a}_k\|_2 \|\boldsymbol{b}_k\|_2\}_{k=1}^{p}$ \citep{xue2022heterogeneous} to improve mediator selection efficiency with moderate sample size.

    Another possible extension is to impose a tensor-based low-rank structure on the coefficient array. As the regression coefficients form a three-dimensional tensor over individuals, time, and mediators, tensor decomposition \citep{bi2021tensors} could provide a more parsimonious representation and enhance information sharing.In this work, we use a fused penalty on latent factors over adjacent time points to promotes temporal smoothness without enforcing a strict tensor structure. Incorporating explicit tensor decomposition remains a promising direction when the true data-generating process exhibits low-rank tensor properties.}

\paragraph{Acknowledgements:}
The authors would like to thank the reviewers, an Associate Editor and the Editor for their constructive comments that improved the quality of this paper. This work is supported by US NSF Grant DMS-2210640, DMS-2515275, and DMS-2515698.

\paragraph{Disclosure statement:}
The authors report there are no competing interests to declare.

	\bibliographystyle{Chicago}
	
	\bibliography{Bibliography-MM-MC}
\end{document}